\documentclass[aps,twocolumn,showkeys,superscriptaddress]{revtex4-1}
\usepackage{graphicx}
\usepackage{amsmath}
\usepackage{amssymb}
\usepackage{longtable}
 \usepackage{dcolumn}
\usepackage{color}
 \usepackage{natbib}

\begin{document}

\title{Role of hydration and intramolecular interactions in the helix-coil transition and helix-helix assembly in a deca-alanine peptide}
\author{Dheeraj S.~Tomar}
\affiliation{Department of Chemical and Biomolecular Engineering, Johns Hopkins University, Baltimore, MD 21218}
\author{Val{\'e}ry Weber}
\affiliation{IBM Research, Zurich, CH-8803 R\"{u}schlikon, Switzerland}
\author{B. Montgomery Pettitt}
\affiliation{Sealy Center for Structural Biology and Molecular Biophysics, Department of Biochemistry and Molecular Biology, University of Texas Medical Branch, Galveston, TX 77555}
\author{D. Asthagiri}\email{Corresponding author:  Dilip.Asthagiri@rice.edu}
\affiliation{Sealy Center for Structural Biology and Molecular Biophysics, Department of Biochemistry and Molecular Biology, University of Texas Medical Branch, Galveston, TX 77555}
\affiliation{Department of Chemical and Biomolecular Engineering, Rice University, Houston, TX 77005}

\date{\today}

\begin{abstract}
For a model deca-alanine peptide the cavity (ideal hydrophobic) contribution to hydration favors the helix state in the coil-to-helix transition and the paired helix bundle in the assembly of two helices.  The energetic contributions of attractive protein-solvent interactions are separated into a short-range part arising from interactions with solvent in the first hydration shell and the remaining long-range part. In the helix-coil transition, short-range attractive protein-solvent interactions outweigh hydrophobic hydration and favor the unfolded coil states. Analysis of enthalpic effects shows that it is the favorable hydration of the peptide backbone that favors the unfolded state. Protein intramolecular interactions favor the helix state and are decisive in folding. In the pairing of two helices, the cavity contribution outweighs short-range attractive protein-water interactions.  However, long-range, protein-solvent attractive interactions can either enhance or reverse this trend depending on the mutual orientation of the helices. In helix-helix assembly, change in enthalpy arising from change in attractive protein-solvent interactions favors  disassembly. In helix pairing as well, favorable protein intramolecular interactions are found to be as important as hydration effects. 

\end{abstract}

 \maketitle

Helices have been termed the ``hydrogen atoms of biomolecular complexity" \cite{zewail}.
In proteins the $\alpha$-helix is a common structural motif and understanding the formation of $\alpha$-helices occupies a pre-eminent place in efforts to understand protein folding.  Using computer simulations and a new approach to free energy calculations \cite{weber:jcp11,Weber:jctc12}, here we revisit this  classic problem and  study two transitions in a model deca-alanine peptide. Mirroring the primary-to-secondary and secondary-to-tertiary transitions in 
protein folding, we study, respectively, the coil-to-helix transition and the pairing of helices to form a helix dimer. Our principal focus is to explicate the hydrophobic and hydrophilic hydration contributions in these transitions at a level that has hitherto been possible only for small molecular solutes. 

There are several reasons to re-examine the fundamental premises about hydration thermodynamics of proteins and the forces driving protein folding.
First, recent experiments and simulations challenge the conventional view that hydrophobic interactions drive protein folding. These studies show that the all-backbone polyglycine, and analogous archetypes of intrinsically disordered peptides,  can undergo a collapse transition in water  \cite{Kiefhaber:pnas06,Crick:pnas06,Tran:2008bk,Hu:2010b,Teufel:2011jmb}. Second, recent computer simulations \cite{Boresch:jpcb09,Boresch:bj13,tomar:bj2013,tomar:jpcb14} reveal important physical and conceptual  
limitations in the group-additive approach that has been the backbone of approaches to understand protein hydration thermodynamics in
experiments (cf.\ Ref.~\onlinecite{Tanford:1970rev,makhatadze:1995}). Indeed the prevailing views of protein hydration thermodynamics \cite{kauzmann:59,Tanford:1962,dill:1990ww} trace back to this group-additive reasoning. Lastly,  while simulations can in principle provide a detailed molecular 
thermodynamics understanding of hydration, with some exceptions \cite{helms:2005fw,pettitt:jacs11}, such studies are scarce for realistic proteins
and polypeptides.  The availability of an approach that alleviates this situation presents an opportunity to revisit and potentially re-appreciate
a classic problem in protein folding. 
 
Earlier studies based on continuum solvent or lattice models have come to differing conclusions about solvent effects in the coil-to-helix transition. 
Some have suggested that hydrophobicity drives the transition \cite{Dill:ps95,Yang:jmb95}, while others have emphasized the role of favorable electrostatics \cite{Avbelj:jmb98}. Experiments  suggest that helix extension is enthalpically driven \cite{Scholtz:pnas91}, as has also been found in computer simulations interpreted within the  Zimm-Bragg or Lifson-Roig formalisms (for example, see 
Refs. \citenum{Garcia:pnasHC02,Gnana:prot05,Best:jpcb09}).  Interestingly, both experiments \cite{makhatadze:jmb04} and computer simulations \cite{Garcia:pnasHC02,Gnana:prot05,Best:jpcb09} show a negative heat capacity upon unfolding, the opposite of what is observed in unfolding of globular proteins \cite{Privalov:1974wo}. The negative heat capacity has been interpreted as arising due to the favorable hydration of the backbone 
upon unfolding  \cite{makhatadze:jmb04}, but a molecular scale description of this signature remained to be sought. 
 
The approach we have developed is based on a quasichemical organization \cite{lrp:apc02,lrp:book,lrp:cpms} of the potential distribution theorem \cite {widom:jpc82}. Using this approach, 
we are able to interrogate the hydration thermodynamics of proteins \cite{weber:jcp11,Weber:jctc12} at a level of resolution that is comparable to those for 
small molecular solutes. Our study on cytochrome C  helped reveal limitations of continuum models of hydration \cite{Weber:jctc12}. 
Subsequent studies have helped illuminate conceptual and physical limitations in the group-additive description of the hydration thermodynamics
of the peptide backbone\cite{tomar:bj2013} and of a hydrophobic side-chain in the context of model peptides \cite{tomar:jpcb14}. On the basis of these new developments, and for the reasons noted above, here we study the hydration thermodynamics in the coil-to-helix and helix-helix assembly in a deca-alanine peptide. 
 
We find that hydrophobic interaction is not consequential in the formation of the helix, but it does play an important  role in helix-dimerization. But hydrophilic hydration is found to play a nontrivial role in the coil-to-helix transition and helix-dimerization.  Indeed hydrophilic hydration can drive the unfolding of the helix and disassembly of the helix dimer. Throughout backbone hydration is found to be the most important component of the overall hydrophilic hydration. For the systems studied here, hydrophilic hydration and protein intramolecular interactions are as important as, if not more important than, hydrophobic effects.

 \section{Theory}
 
The excess chemical potential, $\mu^{\rm ex}$,  is that part of the Gibbs free energy that arises from intermolecular interactions 
and of first interest in understanding the solubility of a solute in a solvent. Here $\mu^{\rm ex}$ is defined relative to the ideal gas at the same density and temperature. To calculate $\mu^{\rm ex}$, we regularize the statistical problem of calculating it from the potential distribution theorem 
$\beta \mu^{\rm ex} =  \ln \langle e^{\beta \varepsilon}\rangle$ \cite{widom:jpc82,lrp:book}, where the averaging $\langle \ldots \rangle$ is over the solute-solvent binding energy ($\varepsilon$) distribution $P(\varepsilon)$. As usual $\beta  = 1/k_{\rm B}T$, with $T$ the temperature and $k_{\rm B}$ the Boltzmann constant.  

\begin{figure}[h!]
\centering
\includegraphics[width=3.25in]{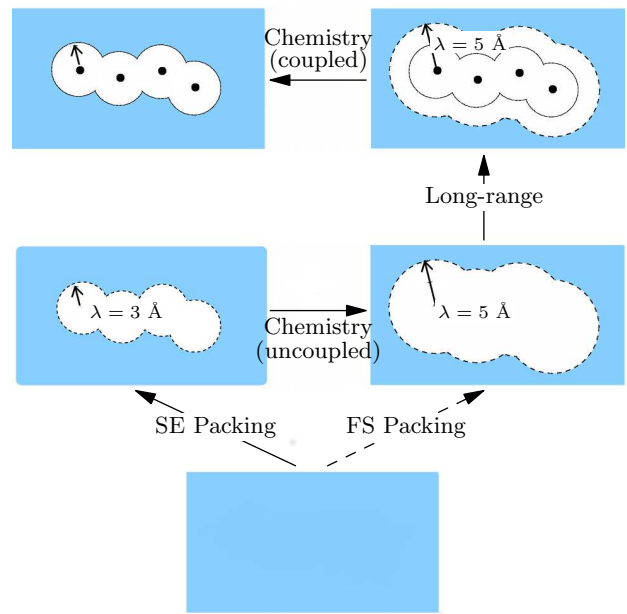}
\caption{Quasichemical organization of the excess chemical potential. The $\lambda = 3$~{\AA} envelope defines the solvent excluded (SE) volume and
the $\lambda = 5$~{\AA} defines the envelope extending to the first hydration shell (FS). For chemistry coupled (uncoupled), the solute-solvent 
interaction is present (absent). In Eq.~\ref{eq:qc} we follow FS-packing to the hydrated solute;  in Eq.~\ref{eq:qc1} we follow SE-packing. Figure adapted from Ref.~\onlinecite{tomar:bj2013} with
permission from Elsevier.}\label{fg:cycle}
\end{figure}

We introduce an auxiliary field $\phi(\lambda; r)$ that moves the solvent  away from the solute to a distance $\lambda$. The distance between the center of the field and the solvent molecule is $r$. For $r > \lambda$,  $\phi = 0$. Since the solvent interface is pushed away from the solute, 
the solute-solvent interaction is tempered and the conditional distribution $P(\varepsilon|\phi)$ is better behaved than $P(\varepsilon)$ \cite{weber:jcp11,Weber:jctc12,tomar:bj2013}. In practice, we adjust the range $\lambda$ such that $P(\varepsilon|\phi)$ is a Gaussian. With the introduction of the field, we have\cite{weber:jcp11,Weber:jctc12,tomar:bj2013}
\begin{eqnarray}
\beta\mu^{\rm ex} = \underbrace{\ln x_0[\phi]}_{\rm local\; chemistry}  + \underbrace{\beta\mu^{\rm ex}[P(\varepsilon|\phi)]}_{\rm long-range} \underbrace{- \ln p_0[\phi]}_{\rm packing}\, , 
\label{eq:qc}
\end{eqnarray}
where $-k_{\rm B}T \ln x_0[\phi(\lambda)]$ is the work done to apply the field in the presence of the solute, $-k_{\rm B}T \ln p_0[\phi(\lambda)]$ is the corresponding quantity in the absence of the solute, and $\beta\mu^{\rm ex}[P(\varepsilon|\phi)]$ is the contribution to the interaction free energy in the presence of the field. Fig.~\ref{fg:cycle} provides a schematic description of 
Eq.~\ref{eq:qc}. 

We apply the field about each heavy atom to carve out the molecular shape in the liquid (Fig.~\ref{fg:cycle}). For convenience we use the same value of $\lambda$ for all the heavy atoms. For $\lambda \geq 5$~{\AA},  $P(\varepsilon|\phi)$ is well-described by a Gaussian.
We thus define $\lambda_{\rm G} = 5$~{\AA}.

The solute also excludes a volume to the solvent and by definition there are no short-range attractive interactions between the
solute and the solvent in this domain. We find that $\ln x_0 \approx 0$ for $\lambda \leq 3.0$~{\AA}, irrespective of the 
conformation of the peptide.  This suggests that the space enclosed by $\lambda_{\rm SE} = 3.0$~{\AA} is excluded to
the solvent and thereby provides a natural definition of molecular extent of the cavity to be used in discussions of the 
cavity (ideal hydrophobic) contribution.  Interestingly, the range between 3 {\AA} to 5 {\AA} corresponds to the first hydration shell for a methyl carbon\cite{asthagiri:jcp2008} and is a conservative description of the first hydration shell of groups containing nitrogen and oxygen heavy atoms.  For simplicity 

Using the $\lambda_{\rm SE}$ and $\lambda_{\rm G}$, we rearrange Eq.~\ref{eq:qc} as 
\begin{eqnarray}
\beta\mu^{\rm ex} & = & \underbrace{\ln \frac{x_0[\phi (\lambda_{\rm G})]}{p_0[\phi(\lambda_{\rm G})]/p_0[\phi(\lambda_{\rm SE})]}}_{\rm revised\; chemistry}   + \underbrace{\beta\mu^{\rm ex}[P(\varepsilon|\phi (\lambda_{G}))]}_{\rm long-range} \nonumber \\ 
& &  \underbrace{-\ln p_0[\phi(\lambda_{\rm SE})]}_{\rm SE\; packing} \, .
\label{eq:qc1}
\end{eqnarray}
Note that  SE packing and  the revised chemistry plus long-range contribution is uniquely defined for the given forcefield. Physically,
the revised chemistry contribution measures the free energy contribution from solute interaction with solvent in the first shell 
relative to a non-interacting (uncoupled) solute (Fig~\ref{fg:cycle}). 

The revised chemistry and long-range contributions describe the role of short-range and long-range attractive protein-solvent interactions in the thermodynamics of hydration. These two components of the hydrophilic contribution occur at different energy (and length) scales.  But 
within commonly used continuum models of electrostatic interactions this distinction is necessarily lost and both the short-range and 
long-range interactions are treated as part of  long-ranged interactions (e.g.~see \cite{dill:1990ww}). From the perspective of such continuum models,  our definition of long-range interaction is more conservative. 

The SE packing contribution measures the hydrophobic hydration of an ideal hydrophobe \cite{Pratt:1992p3019,Pratt:2002p3001}. In theoretical discussions of hydrophobic effects, a cavity with a hard-wall interaction in water is often considered. The packing contribution in our calculation uses
a soft-cavity. This soft-cavity packing estimate is always a lower-bound to the hard-cavity estimate and can be easily corrected to give the latter \cite{weber:jcp11}. We do not pursue those corrections here and instead use the soft-cavity packing result as a measure of primitive hydrophobic effects.  We refer the reader to published papers  \cite{weber:jcp11,Weber:jctc12,tomar:bj2013} for more extensive details about the approach.

\subsection{Helix-Helix PMF}
The above development carries over to the calculation of $W(r)$, the potential of mean force (PMF) to bring two helices a 
distance $r$ apart,  where $r$ is the separation between the helix axis with the axis parallel to each other. 
The PMF 
\begin{eqnarray}
W(r) = W_{solv}(r) + \Delta U(r) \, ,
\label{eq:pmf}
\end{eqnarray}
where $W_{solv}$ is the solvent (or indirect) contribution \cite{asthagiri:jcp2008} and $\Delta U$ is the contribution from direct protein-protein interactions. $W_{solv}(r) =  \mu^{\rm ex}_{\rm dimer} (r) - 2 \mu^{\rm ex}$, where $\mu^{\rm ex}_{\rm dimer}(r)$ is the hydration free energy of the pair of helices (for a given separation and orientation) and $\mu^{\rm ex}$ is the hydration free energy of a monomer helix. 

\subsection{Entropic and enthalpic contributions}

Ignoring the pressure-volume correction and contribution due to a finite isothermal compressibility of water,  the enthalpy of hydration, $h^{\rm ex}$, is given by 
\begin{eqnarray}
h^{\rm ex} = E_{sw} + E_{reorg} \, 
\end{eqnarray}
where $E_{sw}$ is the average peptide-solvent interaction energy and $E_{reorg}$ is the (average) water reorganization energy. 
Additionally ignoring the contribution due to a finite thermal expansivity of water, the entropy of hydration is given by 
\begin{eqnarray}
Ts^{\rm ex} = h^{\rm ex} - \mu^{\rm ex} \, .
\label{eq:sex}
\end{eqnarray}
For calculating $E_{reorg}$, we adapted the hydration-shell-wise calculation described earlier \cite{asthagiri:jcp2008,tomar:jpcb14} (cf.\ Sec.~S.III).

\section{Methods}

The regularization follows previous work \cite{tomar:bj2013}.  The deca-alanine peptide was  modeled with an acetylated (ACE) N-terminus and n-methyl-amide (NME) capped C-terminus.  The extended $\beta$-conformation ($\phi, \psi = -154\pm 12, 149 \pm 9$) was aligned such that the end-to-end vector lay along the diagonal of the simulation cell. We label this coil state as $C_0$. 
The helix was aligned with the helix-axis along the $x$-axis of the cell. The initial structures were energy minimized with weak restraints on the heavy atoms to relieve any strain in the structure. The peptides were solvated in 3500  TIP3P \cite{tip32,tip3mod} water molecules. Version c31 of the CHARMM \cite{charmm} forcefield with correction(cmap) terms for dihedral angles \cite{cmap2},   was used for the peptides.

We sampled unfolded states using the adaptive-bias force (ABF\cite{abf1,abf2}) approach which also additionally provided the
free energy of unfolding the polypeptide in vacuum. From the ABF trajectory, we sampled nine structures with end-to-end distances between terminal carbon atoms ranging between 28~{\AA} and 36~{\AA} in increments of 1~{\AA}. We label the coil states from this unfolding simulation $\{C_1,\ldots,C_9\}$. The $\phi,\psi$ for these unfolded states predominantly populate $\beta$ and PPII regions of the Ramachandran plot.  (Note that, in the strict sense of coil-to-helix transition, the extended $\beta$ is one extreme member in the ensemble of coil states.) In all the hydration free energy calculations, the structures were held rigid. 

The same set-up was used to investigate helix pairing. Additionally, we consider two relative orientations of the helix dipoles, parallel and antiparallel. (Note that  the helix dipoles will be antiparallel in the simplest helix-turn-helix motif.) These arrangements help illuminate the role of long-range protein-solvent interactions in helix-helix complexation. We note that in nature helices almost never align perfectly parallel or antiparallel \cite{scheraga:pnas88}, 
but the hydration effects that are of principal interest here are expected to be insensitive to minor distortions in the relative orientation. 

\section{Results and Discussion}

\subsection{Coil-to-helix transition}

\begin{figure}[h!]
\includegraphics{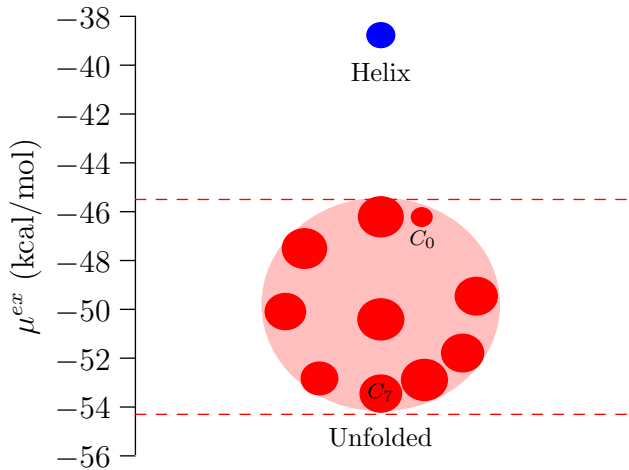}
\caption{Hydration free energies of the helix and $\{C_0,\ldots,C_9\}$ coil states. The horizontal axis has no meaning and 
is used solely to differentiate multiple coil states with similar $\mu^{\rm ex}$ values. The radius of the symbol is equal to twice the
standard error of the mean ($2\sigma$). The $C_0$ state is the smallest circle in the collection of unfolded states; the standard error
is about half compared to the other estimates because  we had 4 times more data for $C_0$ (Sec.~S.I).}\label{fg:unfstates}
\end{figure}

\begin{table*}
\caption{Components of the hydration free energy for the helix and the least favorably ($C_0$) and most favorably ($C_7$) hydrated coil states. For the helix and $C_0$ states,
results with partial charges turned off (indicated by ${\bf Q=0}$) are also noted. $R_g$ is the radius of gyration (relative to the center of mass) and $R_c$ is the end-to-end 
distance between terminal carbon atoms in {\AA}. SASA is the solvent accessible surface area in {\AA}$^2$.  All energy values are in kcal/mol. Standard
error of the mean is given in parenthesis at the $2\sigma$ level.}\label{tb:helixcoil}
\begin{tabular}{|  l  |  c | c | c |  c | c | r | r | c | c | c |}
  \hline
  Conformation & $R_g$ & $R_c$ & SASA & SE  & Revised Chemistry &  Long-range & $\mu^{\rm ex}$ & $E_{reorg}$ & $E_{sw}$ & $Ts^{\rm ex}$ \\
  \hline
  Helix                            & 5.3  & 16.3 &  876.0 & 44.4 (0.4) & $-51.6$ (0.2)&   $-31.6$ (0.2) & $-38.8$ (0.5)  & 75 (6) & $-150.9$ (0.6) & $-37.1$ \\
   Helix (${\bf Q=0}$)  &         &          &             &                   & $-22.4$ (0.3)&   $-18.1$ (0.03) & $3.9$ (0.5)  & 42 (4) & $-59.7$ (0.6) & $-21.6$ \\
  Coil ($C_0$)              &  11.1 &  36.8 & 1260.0 & 58.4 (0.3)  & $-77.1$ (0.3)  & $-27.6$ (0.1)   & $-46.3$ (0.4)  & 92 (6) & $-186.8$ (0.6) & $-48.9$ \\
  Coil (${\bf Q=0}$)     &           &           &              &                       & $-30.9$ (0.4)  & $-24.7$ (0.03) & $2.8$ (0.5) & 50 (8) & $-80.8$ (0.6)  & $-33.6$ \\
  Coil ($C_7$)              & 10.6  &  34.0 & 1249.0 & 58.3 (0.5)  & $-83.4$ (0.3)  &   $-28.3$ (0.4) & $-53.4$ (0.8) & 98 (4) & $-201.4$ (0.6) & $-57.4$ \\ \hline
\end{tabular}
\end{table*}
Figure~\ref{fg:unfstates} collects the results of the hydration free energies of the helix and $\{C_0,\ldots,C_9\}$ coil states; it is clear
that the coil states are better hydrated than the helix. In Table~\ref{tb:helixcoil} we present the free energy components of the helix state,  the least favorably hydrated coil state ($C_0$), and the most favorably hydrated  coil state ($C_7$). 

Before we discuss our results, we first compare our calculations with existing hydration free energy data. As already noted
above, till recently very few all-atom calculations of the hydration free energy of polypeptides with around 10 residues appears
to have been reported. Helms and coworkers\cite{helms:2005fw} have studied blocked-(Ala)$_n$ helix hydration with $n$ up to 9. Extrapolating their results suggests a value of about $-40$~kcal/mol for the deca-alanine helix, consistent with the quasichemical results (Table~\ref{tb:helixcoil}). Kokubo~et~al.\cite{pettitt:jacs11}, computed the van~der~Waals and 
electrostatic contribution to the hydration of a deca-alanine helix and for several coil states. Our hydration free energies based 
on their structures are in fair agreement (data not shown), and the agreement becomes excellent with more extensive sampling in the calculation of the van~der~Waals contribution \cite{Kokubo:jpcb13}. Besides these, the regularization approach
has been tested in studies on water\cite{weber:jcp10a,weber:jcp10b}, ions\cite{beck:jcp13}, and simple peptides \cite{tomar:bj2013},
and the protein cytochrome C\cite{Weber:jctc12}. Further, the quasichemical framework has also been thoroughly documented \cite{lrp:apc02,lrp:book,lrp:cpms}. 

 Turning to the results of this study, Table~\ref{tb:helixcoil} indicates that at least for the coil states considered here, the packing is somewhat insensitive to the peptide  structure. At the scale of the cavities, we expect the packing contributions to scale with surface area \cite{Stillinger:1973tw,chandler:nature05,Ashbaugh:rmp} and the data in the table conforms to this expectation. The free energy per unit SASA is similar to values reported for hard-spheres using scaled-particle theory \cite{ashbaugh:krcpl03,Ashbaugh:rmp} and explicit all-atom simulations \cite{chandler:jcp11,Harris:jcp14,Harris:pnas14}, but is about a factor of ten larger than the surface energy parameter often used in continuum surface-area based model of hydrophobic hydration \cite{sitkoff:jpc94}. Since the solvent-accessible surface areas of $C_0$ and $C_7$ are not very different, despite overall differences in the structure,  we expect the packing contribution to be similar for these states, as found in simulations. The packing contribution favors the helix state by about $-14$~kcal/mol: as expected,  hydrophobic hydration  favors the compact state of the protein. 
 
  The chemistry contribution, however, favors the coil states by between $-31.8$~kcal/mol for the $C_7$ state and $-25.5$~kcal/mol for the $C_0$ state.  Thus,  the local protein-solvent interaction outweighs the packing contribution by between 11~kcal/mol to 18~kcal/mol in favoring the coil state.  Comparing the chemistry contribution for the peptide and its $\bf Q = 0$ analog (Table~\ref{tb:helixcoil}) shows that  the favorable local protein solvent interactions arise primarily from favorable peptide backbone-water interactions, the role of the methyl groups in electrostatic interactions being comparatively negligible. This observation is directly confirmed from the analysis of the solute-solvent interaction contribution to the enthalpy (Table~\ref{tb:helixcoil}). 

Table~\ref{tb:helixcoil} also shows that the entropy of hydration is negative, but here it arises due to attractive solute-solvent interactions. Anticipating our forthcoming article on temperature effects, we note that the entropy calculated using Eq.~\ref{eq:sex} is in agreement within statistical uncertainties with $s^{\rm ex}$ calculated from the temperature derivative of $\mu^{\rm ex}$, as is expected for thermodynamic consistency. 

Results for the $\bf Q=0$ analog  suggests that attractive solute-water dispersion interactions alone can inhibit chain collapse brought about by packing effects (Table~\ref{tb:helixcoil}). The enthalpy of hydration is large negative, despite the positive contribution from solvent reorganization (Table~\ref{tb:helixcoil}), emphasizing the role of attractive dispersion interactions between the $\bf Q=0$ peptide and solvent. These observations emphasize that care is needed in assuming the relevance of the poor solubility of nonpolar solutes in rationalizing the collapse of a polypeptide. Our observation that hydration does not explain the collapse of a nonpolar chain is consistent with the observation of similar behavior in alkanes (cf.\ Ref.~\onlinecite{Ferguson:2009cf} and the reanalysis of data in Ref.~\onlinecite{Athawale:2007fh} presented therein). Interestingly, attractive solute-water interactions also oppose the pairing of the prototypical hydrophobe methane \cite{asthagiri:jcp2008}; emphasizing the importance of solute-water attractions, this effects 
is more pronounced for larger alkanes \cite{BenAmotz:jpcl15}. 

Protein-solvent long-range interactions (cf.\ Fig.~\ref{fg:cycle}) contribute a substantial fraction --- between 50\% to 82\% --- of the net hydration free energy of the peptide. About 90\% of the (favorable) long-range contribution for the $C_0$ and $C_7$ states arises from van~der~Waals interactions, while for the helix it is about 60\%.  As can be expected, the electrostatic contribution for the helix is higher because of the high dipole moment of the helix. (The long-range contributions are  non-negligible because the number of solvent molecules in the outer-shell domain is not small, although individual water-solute interaction energies are typically small.)  The long-range electrostatic and van~der~Waals contributions balance in the coil-to-helix transition resulting in a net free energy change of about $-4.0$~kcal/mol in favor of the helix (Table~\ref{tb:helixcoil}).

The above analysis shows that the hydrophilic contributions to hydration outweigh the hydrophobic driving force and favor the \textit{unfolded} state of the peptide. Experiments show that a coil-to-helix transition in a predominantly alanine-based peptide can occur for a polypeptide chain comprising as few as 13 residues \cite{Spek:1999ts}. We expect the role of hydration in disfavoring the coil-to-helix transition to hold for this slightly larger chain, provided the coil states are such that the backbone remains accessible to solvent. This then suggests that the experimentally observed coil-to-helix transition must be driven by changes in protein intra-molecular interactions, an inference that is in consonance with the suggestion that enthalpy changes driving helix formation \cite{Scholtz:pnas91}, albeit in longer chains. Results in Table~\ref{tb:hexhelixcoil} support this expectation. 

Table~\ref{tb:hexhelixcoil} shows that the favorable hydration of the backbone is lost in coil-to-helix transition ($\Delta E_{bb} > 0$) and this is larger than the
change in enthalpy of solvent reorganization ($\Delta E_{reorg} < 0$) which favors the more compact helical state. It is clear that a favorable change in the internal energy ($\Delta E_{int} < 0$) is necessary to obtain a favorable
change in the net enthalpy $\Delta h^{\rm ex} < 0$. 
\begin{table}[h!]
\caption{Components of the enthalpy change in the coil-to-helix transition for the coil states noted below.  The change in excess enthalpy ($\Delta h^{\rm ex}$)
reveals the role of hydration and the change in internal energy ($\Delta E_{int}$) the role of intra-molecular interactions.  
$\Delta h^{\rm ex}$, is further divided into $\Delta E_{reorg}$, the change in the water reorganization contribution, and $\Delta E_{sw}$, the peptide water interaction contribution. The latter is subdivided into contributions from the backbone-water, $\Delta E_{bb}$, and sidechain-water, $\Delta E_{sc}$, interactions.  Standard error of the mean is given in parenthesis at the $1\sigma$ level. All values are in kcal/mol.}\label{tb:hexhelixcoil}
\begin{tabular}{|  l  |  c | c | c | c | c | }
  \hline
 Coil  & $\Delta E_{reorg}$ &  $\Delta E_{bb}$ & $\Delta E_{sc}$ & $\Delta E_{int}$ & $\Delta h_{\rm total}$  \\  \hline
$C_0$   &  $-17.0 (8.0)$         &  $34.0 (0.4)$       & $1.9 (0.2)$         &  $-43.3$            & $-24.0 (8.0)$ \\
$C_7$   &  $-23.0 (6.0)$         & $51.1 (0.2)$        & $-0.6 (0.2)$        &  $-55.0$            & $-28.0 (6.0)$ \\ \hline
\end{tabular}
\end{table}

On a per-residue basis the net change in enthalpy in the coil-to-helix transition is estimated to be between $-2.4\pm 0.8$~kcal/mol/residue ($C_0$) and $-2.8\pm 0.6$~kcal/mol/residue ($C_7$), with statistical uncertainty reported at the $2\sigma$ level.  A direct comparison of our estimated enthalpy change per residue with experiments is hampered by 
(a) the lack of a rigorous conformational averaging in our calculations, (b) the short length of our polypeptide,  and (c) the fact that in experiments there are residues besides alanine to aid in solubilizing the peptide. Experiments on predominantly alanine peptide (with about 50 residues) \cite{Scholtz:pnas91} suggests a value of $-1.0$~kcal/mol/residue.  Theory suggests that coil states that are less well-hydrated, similar to the $C_0$ state, will dominate the net hydration thermodynamics (cf.\ Ref.~\onlinecite{merchant:jcp09,dixitpd:bj09}) for a corresponding result for ions). These are also the states that appear to have an enthalpy change closer to the experimental result (within statistical uncertainties of the calculation). Forcefield bias can be an issue \cite{Best:bj08}, but  using the recently re-optimized variant (C36 \cite{charmm36}) of the forcefield changes our results by only an additional 5\%. 

\subsection{Helix-helix complexation}

We next consider the free energy of helix association or the potential of mean force between two helices.
\begin{figure}[h!]
\includegraphics{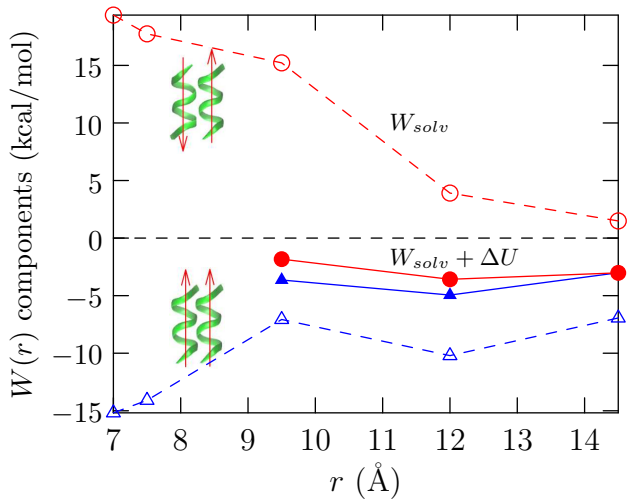}
\caption{Components of the potential of mean force in bringing two helices together. The helices are shown in green and the red arrows
indicate their mutual orientation.  $W_{solv}$ is the solvent contribution (open symbols), and $W_{solv}+\Delta U$ (Eq.~\ref{eq:pmf}) is the net PMF (filled symbols).
For $r \lesssim 8$~{\AA}, there is steric overlap between the helices and $\Delta U$ rises rather sharply. Data including these values of $\Delta U$ are thus not shown.}\label{fg:pmf}
\end{figure}
Fig.~\ref{fg:pmf} shows that hydration opposes the complexation of helices in the antiparallel orientation. Given that the repulsion starts  at a considerable interhelical distance, our choice of helical registration is probably a minor effect.  Our results suggest that hydration  will oppose formation of the helix-turn-helix motif. Interestingly, the direct intramolecular contributions ($\Delta U$) can outweigh the hydration effects to drive complexation. In the antiparallel arrangement favorable $\Delta U$ drives complexation, whereas for the  parallel arrangement, unfavorable $\Delta U$ tempers the favorable hydration effects.  Reminiscent of protein folding free energies, the net free energy of complexation is roughly $-2$~kcal/mol, a small magnitude relative to the large competing hydration and inter-molecular interaction effects. 

\begin{figure*}[ht!]
\includegraphics[scale=0.66]{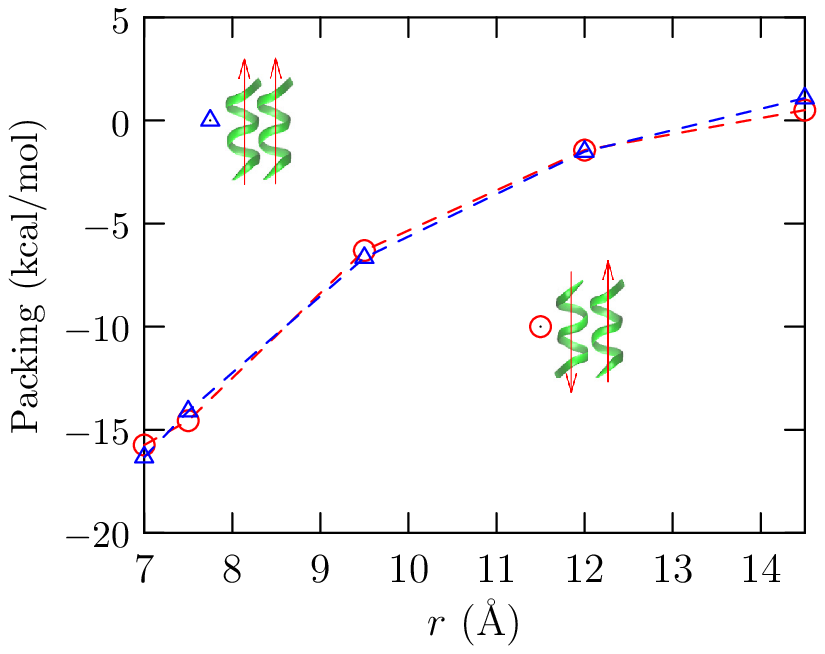}\hspace{3mm}\includegraphics[scale=0.66]{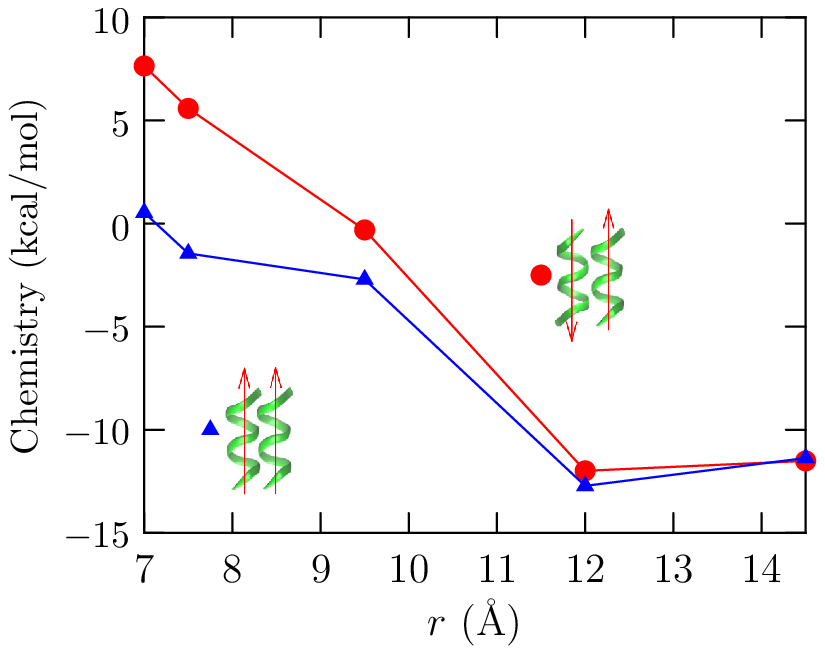}\hspace{3mm}\includegraphics[scale=0.66]{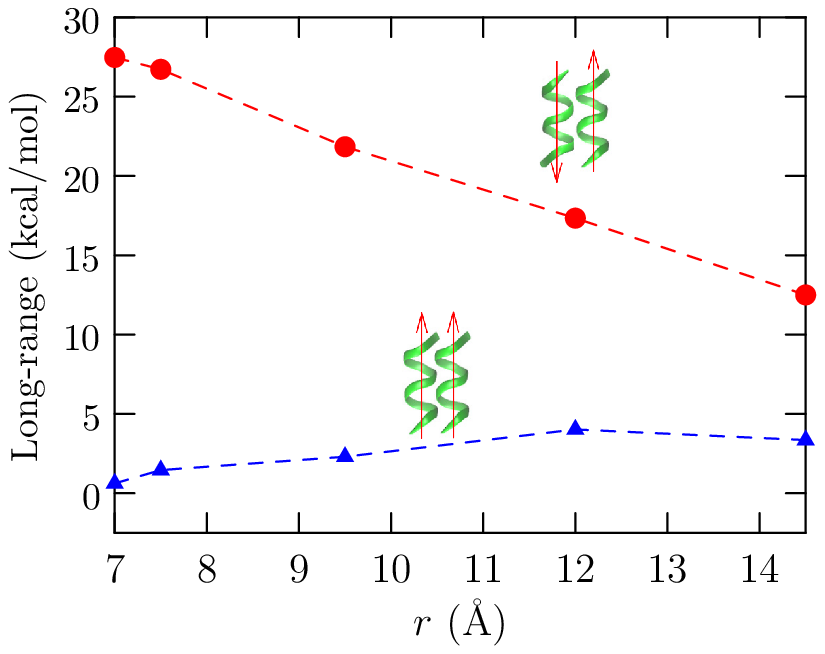}
\caption{The SE packing, revised chemistry, and long-range (Eq.~\ref{eq:qc1}) contributions to the free energy of helix-helix complexation. The data is presented 
relative to two helices infinitely apart. At contact ($r \approx 9.5$~{\AA}) the packing effects outweigh the local chemistry effects, but this trend can easily reversed
by long-range interactions.}\label{fg:pmfcomponents}
\end{figure*}

Figure~\ref{fg:pmfcomponents} shows that primitive hydrophobic effects do drive helix-helix complexation and, in contrast to the coil-to-helix transition, they 
do outweigh the local chemistry contributions. Thus packing (hydrophobic) effects do become important at larger length-scales. 
However, the long-range hydration interactions play an entirely nontrivial role in the complexation process, despite the peptides being net neutral; these
effects can easily outweigh the hydrophobic contribution. The antiparallel arrangement of the helices is strongly disfavored by loss of favorable solute-solvent interactions in hydration, while the inhibition is more modest for the parallel arrangement of helices. 

\begin{figure}[ht!]
\includegraphics{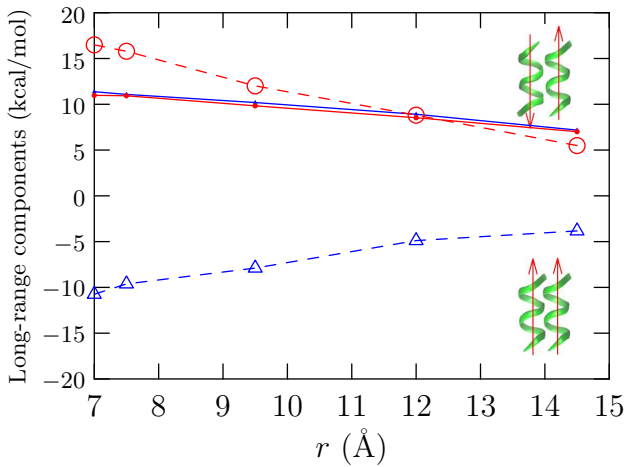}
\caption{The van~der~Waals (solid line) and electrostatic (open symbols, dashed lines) contributions to the free energy. The long-range contribution to free energy is uniquely decomposable \cite{paulaitis:corr10,tomar:bj2013} into electrostatic and van~der~Waals contributions because $P(\varepsilon | \phi) = P(\varepsilon_{vdw} | \phi) \cdot P(\varepsilon_{elec} | \phi)$, where $\varepsilon = \varepsilon_{vdw} + \varepsilon_{elec}$, i.e.\ the individual binding energy distributions are uncorrelated.}\label{fg:longrange}
\end{figure}

Fig.~\ref{fg:longrange} shows that the orientation dependence of the long-range contribution arises solely due to electrostatic interactions. While van~der~Waals interactions between the solute and solvent (outside in the first hydration shell) do not discriminate between the two orientations, its magnitude is non-negligible on the scale of the helix complexation free energy. The parallel arrangement of the helices is favored because of the synergistic effect of the helix macro-dipole; the loss of hydration of the helix dipole also explains the unfavorable contribution for the anti-parallel arrangement. 

Analysis of enthalpic and entropic effects in pairing shows that for both parallel and antiparallel configurations entropic effects favor complexation (Table~\ref{tb:pmfdecomp}), but enthalpic effects do not. However, the characteristics of the change in reorganization and interaction components emphasizes the need for caution in interpreting the entropic driving force in terms of 
changes in water structure. For the antiparallel configuration, the change in the water reorganization energy favors helix pairing, but its effect is negligible for pairing of parallel helices. For both orientations, the loss of hydrophilic backbone-solvent and sidechain-solvent interactions inhibits helix association.  Emphasizing the importance of electrostatic interactions between the backbone and the solvent, the backbone-solvent contributions are sensitive to the orientation of the helices, but the sidechain-solvent contributions are essentially of similar magnitude.
\begin{table*}
\caption{Enthalpic ($h^{\rm ex}$) and entropic contributions ($T\Delta s^{\rm ex}$) to the change in the hydration contribution to the potential of mean force, $W_{solv}$, as the helices are brought from 14.5~{\AA} to 9.5~{\AA}. The change in hydration enthalpy, $\Delta h^{\rm ex}$, is further divided into $\Delta E_{reorg}$, the change in the water reorganization contribution, and $\Delta E_{sw}$, the peptide water interaction contribution. The latter is subdivided into contributions from the backbone-water interactions, $\Delta E_{bb}$, and sidechain-water interactions, $\Delta E_{sc}$. $\Delta X = X(9.5~{\rm {\AA}}) - X (14.5~{\rm {\AA}})$, for all $X$. Standard error of the mean is given in parenthesis at the $1\sigma$ level.
The value of $Ts^{\rm ex} (r = 14.5~{\rm {\AA}})$ is explicitly noted to contrast with the sign of $\Delta E_{reorg}$. All units are kcal/mol.
}\label{tb:pmfdecomp}
\begin{tabular}{|  l  |  c | c | c |  c | c | c | c| }
  \hline
  Conformation & $\Delta E_{bb}$ & $\Delta E_{sc}$ &  $\Delta E_{reorg}$ & $\Delta h^{\rm ex}$ & $\Delta W_{solv}$ & $T\Delta s^{\rm ex}$ & $Ts^{\rm ex} (r=14.5~{\rm {\AA}})$ \\
  \hline
 Antiparallel     & 24.7 (0.5)  & 6.2 (0.1) &  $-7.5$ (5.0) &  $23.4$ (5.0)  & $13.8$ (1.2) &   $9.6 $ & $-70.2$ \\
Parallel             & 4.3 (0.4) & 4.1 (0.1) & $0.8$ (5.0) & $9.2$ (5.0)  & $-0.1$ (1.2)  &   $9.2 $  & $-69.3$ \\ \hline
\end{tabular}
\end{table*}

\section{Conclusions}

We have considered the hydration contributions to both secondary and tertiary protein structure formation by considering the change of a 
solvent-exposed coil to helix  and the association of two such helices in a model decaalanine peptide. The latter idealized model does ignore  the role the loops connecting the helices play in the pairing, but it  is a reasonable starting point to understand solvent effects in tertiary structure formation.

The association of the cavities, the prototypical hydrophobic interaction, tends to favor the compact state of the polypetide and favor both the coil-to-helix transition and helix-helix complexation. But in the coil-to-helix transition, hydrophilic effects (protein-water attractive interactions) easily overwhelm the hydrophobic contribution and favor  unfolding of the peptide. 
 Even for a discharged peptide, essentially a nonpolar chain, attractive solute-solvent dispersion interactions suffice to favor the unfolded state. 

In the pairing of helices,  hydrophobic interactions out weigh the short-range peptide-water interactions in favoring helix complexation. This  occurs at a larger length-scale than the coil-to-helix transition of a single peptide. However, long-range protein-solvent attractive interactions, especially for the antiparallel arrangement of helices,  outweighs the net effect of the packing plus short-range attraction contributions to favor the disassembly of the helices.

In both coil-to-helix transition and helix-pairing, the predominant hydrophilic effects (in our model system) arise from the interaction 
of the backbone with water. This observed importance of the backbone appears to be consistent with recent studies that 
encourage a re-appreciation of the role of the backbone in protein folding (for example, see \onlinecite{Kiefhaber:pnas06,Crick:pnas06,Tran:2008bk,Hu:2010b,Teufel:2011jmb,Bolen:rev08,Rose:pnasbackbone}). 

We find in both the coil-to-helix transition and in the pairing of the helices in the antiparallel orientation,  changes in the intra-molecular energy of the protein are essential in shifting the balance to the folded state. The limitations of the models and forcefield notwithstanding, our study suggests that  in protein folding hydrophilic effects and protein intra-molecular interactions are as important as, if not more important than, hydrophobic effects. 

\begin{acknowledgements}
We thank  Lawrence Pratt and Mike Paulaitis for their critical reading of the manuscript and for helpful discussions. 
We thank Chris Roberts and Bramie Lenhoff for encouraging comments.  We thank Gerhard Hummer and
Mike Pacella for helpful comments on an early version of the manuscript. DA and BMP gratefully acknowledge the financial support of the National Institutes of Health (GM 037657),  the National Science Foundation (CHE-1152876) and the Robert A. Welch Foundation (H-0037).
This research used resources of the National Energy Research Scientific Computing Center, which is supported by the Office of Science of the U.S. Department of Energy under Contract No. DE- AC02-05CH11231.
\end{acknowledgements}


\end{document}